\documentclass[a4paper,fleqn]{cas-dc}
\usepackage[numbers]{natbib}
\usepackage{amssymb}
\usepackage{lineno,hyperref}
\usepackage{graphicx}
\usepackage{xcolor,stfloats}
\usepackage{subfigure}
\usepackage{amsmath}
\usepackage{algorithm}% http://ctan.org/pkg/algorithms
\usepackage{algorithmic}
\usepackage{booktabs} % \toprule, \midrule, \bottomrule
\usepackage{multirow}
\usepackage{graphicx}
\usepackage{xcolor,stfloats}
\usepackage{stfloats}
\usepackage{verbatim}
\usepackage{amscd}
\usepackage{url}
\usepackage{amsfonts}
\usepackage{dcolumn}
\usepackage{tabularx} 
\usepackage{setspace}
\usepackage{bm}
\usepackage{multirow}
\usepackage{textcomp}
\usepackage{threeparttable}
\usepackage{setspace}
\usepackage[figuresright]{rotating}
\usepackage{color}
\usepackage{comment}

\usepackage{times}

\begin{document}
\let\WriteBookmarks\relax
\def\floatpagepagefraction{1}
\def\textpagefraction{.001}

\shorttitle{MIARec: Mutual-influence-aware Heterogeneous Network Embedding for Scientific Paper Recommendation}

\shortauthors{Wenjin Xie et~al.}

\title [mode = title]{MIARec: Mutual-influence-aware Heterogeneous Network Embedding for Scientific Paper Recommendation}

\author[inst1]{Wenjin Xie}[orcid=0000-0002-1438-8443]
\ead{xiewenjin@email.swu.edu.cn}

\author[inst1,inst2]{Tao Jia}[orcid=0000-0002-2337-2857]
\ead{tjia@swu.edu.cn}
\cormark[1]

\affiliation[inst1]{organization={College of Computer and Information Science, Southwest university},%Department and Organization
            %addressline={Tiansheng Road No.2}, 
            postcode={400037}, 
            state={Chongqing},
            country={China}}
\affiliation[inst2]{organization={College of Computer and Information Science, Chongqing normal university},%Department and Organization
            %addressline={Tiansheng Road No.2}, 
            postcode={401331}, 
            state={Chongqing},
            country={China}}

\cortext[cor1]{Corresponding author}

\begin{abstract}
With the rapid expansion of scientific literature, scholars increasingly demand precise and high-quality paper recommendations. Among various recommendation methodologies, graph-based approaches have garnered attention by effectively exploiting the structural characteristics inherent in scholarly networks. However, these methods often overlook the asymmetric academic influence that is prevalent in scholarly networks when learning graph representations. To address this limitation, this study proposes the Mutual-Influence-Aware Recommendation (MIARec) model, which employs a gravity-based approach to measure the mutual academic influence between scholars and incorporates this influence into the feature aggregation process during message propagation in graph representation learning. Additionally, the model utilizes a multi-channel aggregation method to capture both individual embeddings of distinct single relational sub-networks and their interdependent embeddings, thereby enabling a more comprehensive understanding of the heterogeneous scholarly network. Extensive experiments conducted on real-world datasets demonstrate that the MIARec model outperforms baseline models across three primary evaluation metrics, indicating its effectiveness in scientific paper recommendation tasks.
\end{abstract}

%\centering
%\includegraphics[width=0.9\textwidth]{framework.pdf}

%\item Research highlights item 1
%\item Research highlights item 2
%\item Research highlights item 3


\maketitle


\section{Introduction}

When engaged in scientific research, scholars frequently require a significant amount of reference literature to assist them in recent advancements in their fields, expanding their knowledge, and inspiring new ideas. With the exponential growth in the volume of scientific literature, scholars are confronted with a common challenge: how to quickly and accurately identify valuable scientific papers from massive materials. The personalized scientific paper recommendation systems address this challenge by offering individually tailored suggestions, moving beyond basic popularity or recency, thus becoming a critical area of research endeavor \citep{li2019review, kreutz2022scientific}.

Traditional paper recommendation methods can be generally classified into content-based (CB) methods and collaborative-filtering-based (CF) methods. CB methods often infer a scholar's preferences based on the content of their own papers or related papers and make recommendations accordingly \citep{basu1998recommendation, lu2007node}. In contrast, CF methods operate inversely, recommending to a scholar based on the preferences of other scholars whose past behaviors are similar to those of the target scholar \citep{charlin2014leveraging}. Although these methods have achieved certain successes in scientific paper recommendation tasks, they also face several challenges. Firstly, in the case of cold start and data sparsity, which are common issues in recommendation systems, these methods may struggle to provide accurate recommendations due to the lack of historical data, like user profile information or behavior, for new users. Secondly, the above-mentioned methods focus on the relations between scholars and papers, neglecting the worthwhile relations among scholars. In fact, scholars can be linked through various relationships, including collaborations, publishing in the same venues, and sharing research interests. Thus, the graph-based (GB) recommendation methods are proposed to capture the structural information of scholars within scientific networks, and assess node proximity to facilitate recommendations \citep{xia2016scientific, amami2017graph, pan2015academic}. 

With the rise of graph neural network technology, GNN-based methods for recommending scientific papers have gained popularity due to their capacity to process structured data, such as scholar collaboration graphs and citation graphs \citep{huang2015neural, wang2020deep, hao2021paper, liu2024evolving}. However, there are still limitations to GNN-based methods. On the one hand, the structural heterogeneity of scientific networks is difficult to capture through these methods. Different from the networks whose nodes and edges are of a single type, named homogeneous networks, such as power grids and online social networks, the networks with multiple types of nodes and edges are called heterogeneous networks. The scientific network is a typical heterogeneous network characterized by diverse types of nodes (e.g., authors, papers, venues) and various types of relationships (e.g., co-authorship, citation, publication). Therefore, homogeneous network representation learning methods alone cannot accurately and completely capture the structural features inherent in the heterogeneous network\citep{metapath2vec, han}. In this paper, we focus on the heterogeneous network composed of scholars and their multi-relational connections. By employing the graph learning approach, the structural relationships among scholars are obtained, thereby enhancing the accuracy of recommendations. 

On the other hand, GNN approaches for learning node embeddings typically aggregate feature information from the neighbors. In the aggregation process, the weights associated with adjacent nodes are indicative of the influence exerted by neighboring nodes on the target node. This influence can be categorized into two distinct methodologies: symmetric normalization \citep{gcn, graphsage} and deep attention mechanisms \citep{gat}. However, the approaches often fail to account for the asymmetrical influence of edge weights. This asymmetry emerges in social network contexts, where individuals of high stature, such as celebrities, have a significantly greater influence over their followers than the influence those followers exert in return \citep{fraser2002media, bush2004sports, moraes2019celebrity}. An analogous phenomenon can be observed within scientific networks, where prominent scholars with high academic prestige often yield a huge influence over their peers \citep{kang2016scientific, caulfield2016science}, evident in metrics like paper citations and topic selection \citep{fu2023mutual, amjad2016muice}. To address the issue of asymmetric influence, we introduce the mutual-influence-aware mechanism that reflects academic influence during node feature aggregation. This enhances the sensitivity of the feature information propagation process in network representation learning, leading to more accurate structural representations of scholars and improved recommendation performance.

Based on the above discussion, in this paper, we propose the scientific paper recommendation model with a mutual-influence-aware multi-channel aggregating heterogeneous network embedding method, named MIARec. The model not only learns the embeddings of scholars within each single-relational network extracted from heterogeneous scientific networks, but also directly learns the interdependent embedding representation of all the sub-networks. In the embedding process, the representation of a node is learned by the convolution of neighbor features where the weight relies on the mutual influence between nodes. We further utilize the deep attention mechanism to automatically learn the importance weights for different embeddings in all channels, so as to adaptively aggregate them. Combined with the representation of paper text features, the model predicts the most relevant papers for the target scholar as recommendations by assessing the similarity between those representation vectors.

We summarize our main contributions as follows:

\begin{enumerate}%[label=(\arabic*)]
\item We propose a multi-channel heterogeneous network representation learning framework for recommendation that models single-relational subgraphs independently, enables joint learning through parameter sharing in an interdependent embedding channel, and aggregates embeddings via attention.

\item Considering the asymmetry and imbalance in academic influence among scholars, we propose a GNN-based approach that incorporates mutual influence awareness to more accurately learn the structural representations of scholars within the academic networks.

\item Extensive experiments on the real-world academic datasets have been conducted. The model comparison results reveal that the proposed model outperforms baseline models by an average of 0.78\%, 2.60\% and 3.17\% across three main metrics on DBLP datasets and 1.28\%, 1.72\% and 2.84\% on ACM datasets. The ablation study confirms the effectiveness of each module in the model.
\end{enumerate}

The rest of the paper is organized as follows. In Section 2 we introduce the literature review related to this work. Section 3 presents the definition of the heterogeneous network and the scientific paper recommendation task. In Section 4, we develop our recommendation model (MIARec) including academic network embedding and recommendation method. We report and analyze experimental results in Section 5. Finally, Section 6 draws a conclusion with future work.

\section{Related works}

\subsection{Heterogeneous network embedding}
Compared to homogeneous networks, heterogeneous networks have attracted extensive scholars' attention due to the rich structural and semantic information brought by their diverse nodes and edges \citep{shi2016survey}. \cite{metapath2vec} transfers embedding methods from homogeneous networks to heterogeneous networks for the first time by proposing Metapath2vec. Meta-path depicts the fixed semantic relationships between different types of nodes. They performed a meta-path-based random walk and used skip-gram to embed meta-path generated from heterogeneous graphs. \cite{han} improves Metapath2vec and introduces the node-level attention to aggregate the neighbors' embeddings to the node and semantic-level attention to aggregate embeddings of different semantic meta-paths. \cite{fu2020magnn} proposes MAGNN, which not only considers the aggregation of semantic nodes within meta paths in the heterogeneous graph but also incorporates aggregation between multiple meta paths. \cite{rgcn} proposes R-GCNs, treating each relational context within a heterogeneous network as some homogeneous subgraphs. Node embeddings are learned in independent subgraphs and aggregated to obtain the whole graph embedding. 

As the scientific network is one of the most typical representatives of heterogeneous networks, it has received widespread attention from many scholars. Extensive works focus on some practical issues in scientific networks. Among them, the research methods based on heterogeneous network embedding have been widely adopted due to their ability to accurately obtain implicit relationships between nodes, thereby helping to solve downstream tasks, such as the collaborator recommendation task and author name disambiguation task.
\cite{liu2018context} jointly embeds authors and article topics in an author-topic heterogeneous network. By combining the contextual information of the collaboration network with the semantic information of the article topic, the embeddings of the author-topic network are learned to recommend collaborators. \cite{yang2020hnrwalker} designs the improved random walk algorithm with dynamic transition probability and a new rule for selecting candidates to obtain a better embedding for the author-institution network. \cite{liu2023hnerec} extracts nodes in different pre-defined meta-paths and utilizes the skip-gram model to embed the nodes into vectors. The final collaborator recommendation list is based on the similarity between the corresponding node vectors. Scholars also address the author name disambiguation problem using network embedding approaches. \cite{qiao2019unsupervised} uses Doc2vec to get the original node embedding and uses a heterogeneous skip-gram method to learn the publication representations, then clusters the representations to determine the assignment of publications. \cite{xie2022author} proposes a meta-path level attention to learn the importance of heterogeneous relations and combines the embedding of each relation graph for disambiguating the author name of publications. Due to the success of network embedding in these tasks, scholars have paid attention to solving the paper recommendation task based on network embedding.

\subsection{Scientific paper recommendation}
Scientific paper recommendation methods can be classified into four categories: content-based, collaborative-filtering-based, graph-based, and hybrid method \citep{bai2019scientific, kreutz2022scientific}.

In content-based (CB) methods, a scholar profile is first created, incorporating their publications and corresponding papers. Next, feature vectors for both the scholar and papers are generated using the TF-IDF model \citep{bollacker1998citeseer, jomsri2010framework}, Latent Dirichlet Allocation (LDA) \citep{amami2016lda}, or the deep learning method \citep{huang2015neural}. Finally, the recommendation list is produced by calculating the similarity between the scholar and paper feature vectors. \cite{zhao2016paper} builds the concept map based on the paper topics and their relationships, then recommends papers by identifying the shortest path on that map to bridge the background knowledge and research target of the scholars. 
The collaborative-filtering-based (CF) methods refer to the preferences of a user's neighbors to make paper recommendations \citep{wang2011collaborative, wang2014relational}. In contrast to CB, the CF models are very effective in recommending relevant papers when content information is not available. \cite{yang2014recommendation} extracts the latent correlation underlying the relations among users and uses a joint CF model to recommend relevant items for users. \cite{bansal2016ask} introduces a model that generates collaborative filtering paper predictions by employing a gated recurrent unit (GRU) network. The model utilizes user friends’ ratings along with the content of papers to make recommendations. The CB and CF methods both encounter the cold-start problem when the user profile information is absent. Besides, Both methods lack the use of structural information and semantic relations within the heterogeneous scientific network, resulting in their inability to capture meaningful semantics and generate relevant recommendations.

With the valid development of deep graph learning, graph-based (GB) methods are commonly employed to leverage the structural information, such as scholar collaboration and paper citation relationships. \cite{huang2015neural} creates citation context recommendations using semantic representations of citation contexts and relevant papers, employing a multi-layer neural network to estimate the probability of citing a paper based on the context. Similarly, \cite{yang2018lstm} proposes PCCR, utilizing LSTM for learning citation context and paper representations through context and paper encoders, respectively. The model identifies top-k citations for a given context by calculating cosine similarity between the context embeddings and candidate papers. \cite{wang2020deep} uses author and citation information with a BiGRU network for context-aware recommendations. In contrast, another category of GB approaches focuses on learning the embedded representations of papers within the scientific network and recommending ones to scholars that are similar to those they have authored. p-CNN leverages CNN to assess relevance between citation contexts and papers, employing a discriminative training strategy for parameter learning and recommendation generation \citep{yin2017personalized}. Likewise, POLAR introduces an attention-based CNN model for paper recommendations, utilizing an attention matrix to capture both local and global weights of salient factors and words \citep{du2019polar}. EKGE constructs a scholarly knowledge graph based on paper information and proposes a dynamic learning approach to obtain representations of papers within this graph, leveraging these representations for recommendation purposes \cite{liu2024evolving}.

The graph-based methods are also combined with other recommendation techniques to enhance the performance. \cite{ma2019personalized} builds the scholar and paper profiles based on the content and embeds the scientific network using a meta-path-based model for recommendation. \cite{hao2021paper} explores the periodic academic interests of scholars based on their past products, and combines the structural information in scientific networks to obtain the similarity between scholars and papers for recommendation. \cite{ali2022spr} proposes an SPR-SMN model that learns both the context-preserving paper content representations and node embedding using semantic relations and long-range dependencies in the network. \cite{zhang2022citation} extracts the relationship among papers by network embedding and learns the paper representations via the content, then combines the representations and makes recommendations based on the link prediction approach.
In this paper, we also take the graph-based approach. The structural representation of heterogeneous scholar networks is extracted using an embedding method that captures mutual academic influence among scholars. Combined with the textual features of papers, this approach yields scientific paper recommendation results.

\section{Preliminary}

\textbf{Definition 1 Heterogeneous network.} 
A heterogeneous network is a graph $G=\{N,E,\mathcal{N},\mathcal{R}$ where each node $n \in N$ is mapping to a node type with the mapping functions $\phi(n): N \rightarrow \mathcal{N}$ and each edge $e \in E$ is mapping to an edge type with $\phi(e): E \rightarrow \mathcal{R}$. In a heterogeneous network, $|\mathcal{N}|+|\mathcal{R}|>2$.

\textbf{Definition 2 Scientific paper recommendation.}
Given a scholar set $\mathcal{S}=\{n_i|i \in |N^s|\}$, a scientific paper set $\mathcal{P}=\{p_i|i \in |N^p|\}$ and a heterogeneous academic network $G$, the scientific paper recommendation task aims to learn a prediction function: $f: \mathcal{S}\times\mathcal{P}\times G \rightarrow \mathbb{R}$, that allows us to predict the relevance score $\hat{Y}_{ij}$ of each uninteracted scientific paper $p_j$ for a target scholar $n_i$, thereby generating a ranked list of papers $\hat{P}_{n_i}=\{\hat{p}_m|m=0,1,2...\}$, where papers with higher relevance scores are positioned at the top of the list. 

\section{Methodology}
In this study, we propose the MIARec model for scientific paper recommendation, which learns representations of scholars within a heterogeneous academic network and derives representations of scientific papers from their content. These two types of representations are integrated for paper recommendations. The framework of the proposed model is demonstrated in Fig. 1.

\begin{figure*}[htb!]
	\centering
	\includegraphics[width=1\textwidth]{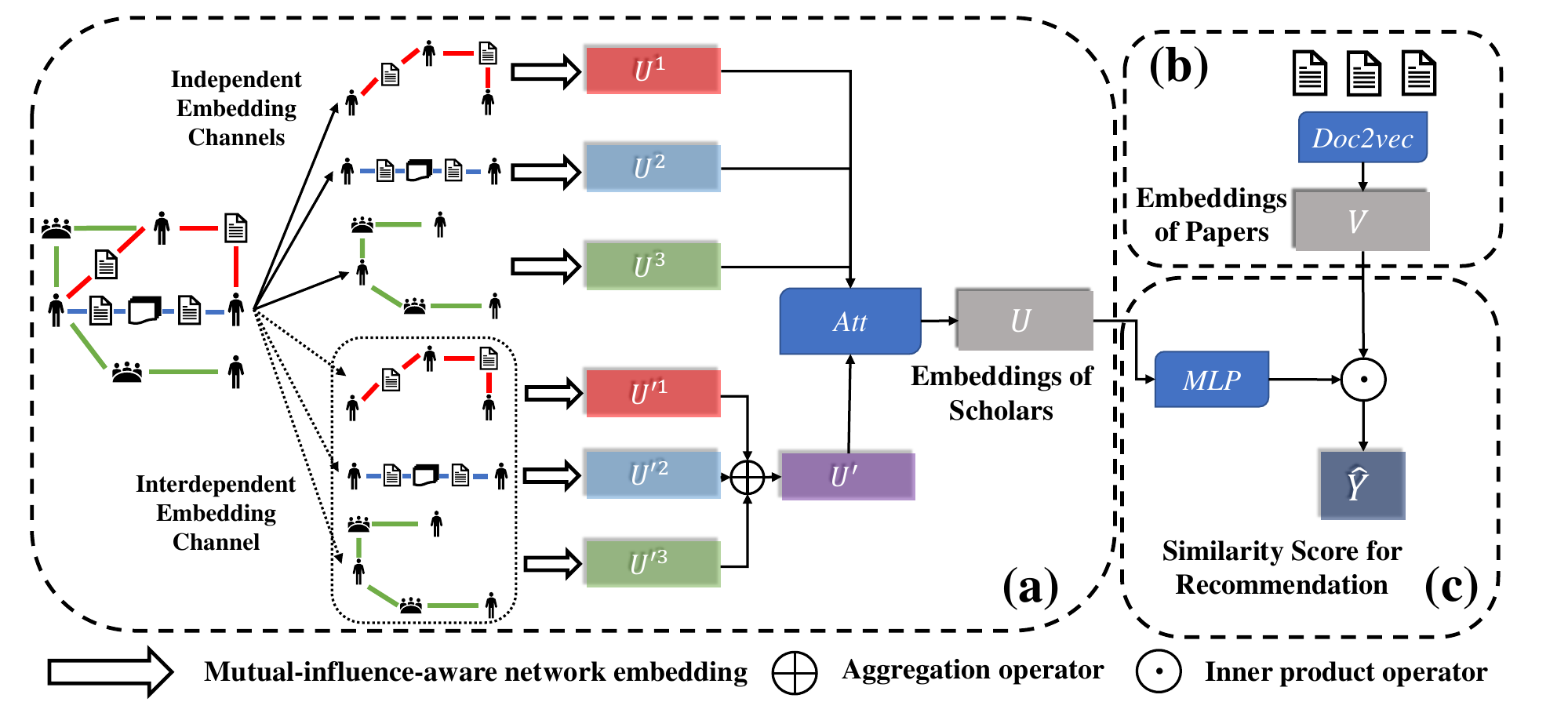}
	\caption{The framework of proposed MIARec model. \textbf{(a)} The scholar representation learning part. The heterogeneous network is decomposed into single-relational sub-networks, each embedded in an independent channel via a mutual-influence-aware embedding module. These sub-networks share embedding parameters within a common interdependent embedding channel and are then aggregated. An attention module fuses these aggregated embeddings into a unified representation for scholars $U$. \textbf{(b)} The paper representation learning part. Scientific papers' embeddings are learned through Doc2Vec. \textbf{(c)} The recommendation prediction part. The model employs a multi-layer perceptron (MLP) to align the two types of embedding representations and performs recommendations based on their inner product similarity.\label{fig:framework}}
\end{figure*}\noindent

\subsection{Mutual-influence-aware scholar embedding}

When dealing with heterogeneous scholar networks with multiple types of relationships, existing methods typically rely on meta-paths to separately extract information for each specific type of relationship or partition the network into sub-networks that contain only single-type relationships \citep{metapath2vec, han}. The underlying principle of these approaches is to learn associations among users within each scholarly relationship individually. However, research has pointed out limitations in this approach, as it focuses primarily on embeddings within individual networks neglecting the latent connections between different relationship types \citep{rgcn}. Therefore, in this study, we not only continue the approach of existing frameworks by employing multiple independent embedding channels to separately learn representations for single-relation sub-network, but also propose an interdependent embedding channel that utilizes a parameter-sharing strategy to enable representation learning across distinct sub-networks, as studies have indicated that parameter sharing can reveal deeper connections among embeddings, thus providing a more integrated perspective of relational structures \citep{wang2020gcn, rgcn}. Finally, all the embeddings are aggregated using the attention mechanism.

\subsubsection{Single-relational networks embedding}

For the author relationships $r_1,r_2,...,r_k$ in a scientific network $G=\{N,E,\mathcal{N},\mathcal{R}\}$, we extract the corresponding single-relational sub-networks $G^1,G^2,...,G^k$ in which all the nodes represent the scholars in the network and the edges are of the same attribute representing a relationship between scholars. In this work, we use three relations: collaboration, co-topic (having at least three identical keywords), co-venue (attending the same venue). In the single-relational sub-network $G^r$, the number of node type and edge type should both be $\mathcal{N}=1, \mathcal{R}=1$, so we simply denote it as $G^r=(N^r,E^r)$.

\paragraph{Independent embedding}
For each single-relational sub-network $G^r$, we feed it to an independent channel to obtain its specific representation embedding. In scientific networks, scholars with connected relationships often exhibit certain similarities. For instance, scholars engaged in similar research domains are likely to have collaboration and citation relationships. Leveraging this characteristic, GCN-like approaches are well-suited for the representation learning tasks, whose underlying principle involves utilizing the feature information of neighboring nodes in the network to obtain the representation of a given node \citep{gcn, zhang2019graph, gat, graphsage}. Among them, the GraphSAGE model \citep{graphsage} is proved to be more suitable than the original GCN for the scientific paper recommendation \citep{santosa2024s3par}, for its ability to handle large-scale graph data by employing a sampling method that reduces graph size without compromising model performance. Inspired by GraphSAGE, the independent embedding module consists of two key steps: sampling and aggregation. During the sampling phase, a fixed number of neighboring nodes are randomly selected for message aggregation, reducing computational complexity and memory usage while retaining local node information. For aggregation, the embedding of the target node is obtained by combining its eigenvectors from the previous layer with the aggregated vectors of its neighboring nodes. The aggregated vectors of the sampled neighboring nodes are derived as:
\begin{equation}
AGG^{(l)}(SNi)=\sigma\left(\frac{1}{SN_i}\sum_{j \in SN_i} \frac{M_{ij}}{\sqrt{\left|AN_i\right|} \sqrt{\left|AN_j\right|}} u_j^{(l)}\right),
\label{eq:conv}
\end{equation}
where $u_i^{(l)}$ and $u_i^{(l+1)}$ individually represent the vector representations of $n_i$ in the $l$-th convolutional layer and the $(l+1)$-th layer; $SN_i$ represents the set of the sampled neighbors of $n_i$; $AN_i$ represents the set of all the neighbors of $n_i$; $M_{ij}$ is the factor measuring the mutual influence of $n_i$ and $n_j$.

When $M_{ij}=1$ in Equation \ref{eq:conv}, it represents a symmetric normalization term consistent with standard GCN design \citep{gcn}. This indicates that feature aggregation from neighboring nodes is generally performed in a symmetric and averaged way, overlooking the asymmetric mutual influence between scholars. When $M_{ij}$ is the attention weight on the edge between $n_i$ and $n_j$, it follows the GAT framework to capture the asymmetrical effects. However, the reliance on an attention mechanism can lead to unstable learning outcomes and often lacks transparency in the decision-making process. In this paper, we design a normalized mutual-influence-aware coefficient:
\begin{equation}
M_{ij} = \frac{exp(g_{ij})}{\sum_{m \in AN_i} exp(g_{im})},
\end{equation}
where $g_{ij}$ represents the academic mutual influence factor between scholar $n_i$ and $n_j$. In the domain of network science, the gravity model is frequently employed to quantify the node influence \citep{li2019identifying, yang2021improved, xu2024cagm}. Thus, we adopted this approach, utilizing the gravity model to quantify the academic influence between scholars, as detailed below:
\begin{equation}
F_{ij} = G \frac{m_im_j}{{r_{ij}}^2},
\end{equation}
where $G$ is the gravitational constant; $m_i$ is set as the citation count of scholar $n_i$; $r_{ij}$ denotes the academic distance between $n_i$ and $n_j$. This distance is defined as the reciprocal of their collaboration count, showing that more frequent collaborations result in a shorter academic distance. Nevertheless, while the gravity model can effectively differentiate the varying degrees of ``academic gravity" exerted by different neighboring scholars, it fails to capture the asymmetry in influence between a pair of scholars. Therefore, to accurately measure the extent of influence that scholar $n_j$ has on scholar $n_i$, we normalize this ``academic gravity" by dividing it by the ``mass" of $n_i$ for this asymmetry and obtain the mutual influence factor $g_ij$:
\begin{equation}
g_{ij} = G \frac{m_j}{{r_{ij}}^2}.
\end{equation}

Thus, in iterative sampling and aggregation, the embedding representation of node $n_i$ in the last layer $L$ is obtained by concatenating its own embedding in $(L-1)$-th layer and the aggregation embedding in $L$-th layer:
\begin{equation}
    u^{(L)}_i = \sigma(W^{(L)}\cdot CONCAT(u_i^{(L-1)},AGG^{(L)}(SN_i))),
\end{equation}
where $W^L$ is a learnable weight matrix that processes the concatenation embedding through a single-layer neural network; $\sigma(\cdot)$ is the activation function which is typically specified as the Rectified Linear Unit (ReLU) function. Combining the final embeddings of all nodes $\{u^L_i, \forall i \in N^{r}\}$, the independent embedding of the single-relational sub-network $U^r,r=1,...,k$ can be obtained in this specific channel.

\paragraph{Interdependent embedding}

To capture the latent correlations among sub-networks, they are all sent to an interdependent channel where the embedding layer parameters are shared.
Analogous to those within the independent channels, the embedding of a node $n_i$ in the $l$-th layer of the interdependent channel can be derived as follows:
\begin{equation}
    u_i^{'(L)} = \sigma(W^{'(L)} CONCAT(u_i^{'(L-1)},AGG^{'(L)}(SN_i)).
\end{equation}
The distinction lies in the fact that the parameters within the weight matrix $W^{'(L)}$ in this context are shared across the embedding representations of all sub-networks, thereby facilitating these parameters to be learned collectively within the interdependent channel. Utilizing this strategy, latent shared features across distinct subnetworks can be extracted. Combining all the interdependent embeddings of sub-networks $\{U^{'r},r=1,...,k\}$, we can obtain an interdependent embedding of the full heterogeneous network:
\begin{equation}
    U^{'} = \frac{1}{k}\sum_{r=1}^{k}U^{'r}.
\end{equation}

\subsubsection{Multi-channel aggregating} %
For single-relational sub-networks with $G^1,G^2,...,G^k$ varying semantics, we learn independent node embeddings $U^{1},U^{2},...,U^{k}$ specific to each channel, and an interdependent embedding $U^{'}$ capturing relationships across all sub-networks. To address the differing impact of each sub-network's semantics on the paper recommendation task, we employ an attention mechanism to automatically learn and fuse the importance of different sub-network representations. The importance of each embedding is calculated as:
\begin{equation}
\alpha^1, \alpha^{2},...,\alpha^{k}, \alpha^{'}=att(U^{1}, U^{2},...,U^{k},U^{'}),
\end{equation}
where $\alpha^r \in \mathbb{R}^{n\times 1}$ denotes the attention weight of $n$ nodes with each embedding and $att$ denotes the attention module using the deep neural network. 

For the scholar $n_i$ in the network, we transform its embedding $u_i^r$ in the network $G^r$ through a nonlinear transformation, and use one shared attention vector $q$ to get the attention weight $\omega_i^r$ as follows:
\begin{equation}
\omega_i^r=q^T \cdot \tanh \left(W^a \cdot\left(u_i^r\right)^T+b\right),
\end{equation}
where $W^a$ is the weight matrix and $b$ is the bias vector. So the attention weight $\omega_r^a$ can be normalized by the $softmax$ function to get the final attention value:
\begin{equation}
    \alpha_i^r = softmax(\omega_i^r) = \frac{exp(\omega_i^r)}{\sum_{j=1}^{k}exp(\omega_{i}^j) + exp(\omega^{'}_i)},
\end{equation}
where $\omega_i^r$ stands for the attention weight of the node $n_i$ embedding in each sub-network and $\omega^{'}_i$ is the attention weight of its interdependent embedding. Here larger $\alpha_i^r$ implies the corresponding embedding $u_i^r$ is more important. Thus for all the nodes, the important weights are learned as the diagonal matrix $\alpha^r = diag[ \alpha_i^r ]$. Finally, we can combine these network embeddings together to obtain the total embedding $U_t$ at time $t$:
\begin{equation}
    U = \sum_{r=1}^{k}\alpha^r \cdot U^r + \alpha^{'} \cdot U^{'}. \label{eq:zt}
\end{equation}

\subsection{Recommendation model and optimization}
To enhance the accuracy and relevance of personalized paper recommendations, it’s important to consider the consistency characteristics of scholars' preferences. Therefore, we should assess both the structured similarity between scholars and the semantic similarity at the paper's textual level. In this study, we utilize the established Doc2Vec model \citep{le2014distributed} to generate semantic embedding vectors for papers, denoted as $V \in \mathbb{R}^{|V|}$. To integrate these with the scholar representation embeddings from the previous module, $U \in \mathbb{R}^{|U|}$, We apply an MLP layer to align scholars' representations with those of the papers:  $U^a=\sigma(W^aU+b^a)$, where $W^a \in \mathbb{R}^{|V| \times |U|}$ is the learnable linear variation matrix and $b^a \in \mathbb{R}^{|V|}$ is the bias vector.

In the training process, we address this implicit feedback problem in scientific paper recommendation by employing the Bayesian Personalized Ranking (BPR) loss for training, a method tailored for user implicit feedback and widely utilized in recommendation systems. It posits that observed interactions, indicative of user preferences, should yield higher prediction scores than unobserved ones. The loss function is formulated as follows:

\begin{equation}
    \mathcal{L}_{\text{BPR}} = - \sum_{(i,j,k) \in \mathcal{D}} \log \sigma(\hat{Y}_{ij} - \hat{Y}_{ik}) + \lambda \left\| \theta \right\|_2^2
\end{equation}
where $\mathcal{D}$ means all scholar-paper pairs in the academic network; the pair $(i,j)$ is the positive sample, indicating that paper $j$ is relevant to scholar $i$, while the pair $(i,k)$ is the negative sample, indicating that scholar $i$ does not interact with paper $k$; $\lambda$ is the regularization weight and $\theta$ stands for the model parameters. The similarity score between pair $(n_i,p_j)$ is calculated as the inner product of the transpose of the aligned scholar embeddings $u_i^{a}$ and the paper embedding $v_j$:
\begin{equation}
\hat{Y}_{ij} = \text{sim}(u_i^{a},v_j) = u_i^{aT} \cdot v_j .
\end{equation}

\section{Experiments}

\subsection{Datasets}

We use the datasets of Aminer DBLP-V12 and ACM-V10 \footnote{https://www.aminer.cn/citation.} \citep{tang2008arnetminer}. The DBLP dataset contains 4,894,081 scientific papers and 45,564,149 citations and the ACM dataset contains 2,579,904 scientific papers and 17,254,368 citations. The information it provides includes papers’ titles, abstracts, authors, venues, keywords, and citation relations.

The dataset's composition follows prior works \citep{jiang2023taprec, wang2023marec, zhang2025mkcrec}, in which the training and testing sets are generated using a leave-one-out strategy. We collect all papers cited by the target scholar's publications as positive samples. Papers cited by the scholar's latest paper form the test set, with the remaining citations used for training. A certain number of irrelevant papers are added to the test set as negative samples, ensuring a positive-to-negative sample ratio of 1:3 in the test set.

\subsection{Baselines}
To verify the effectiveness of our MIARec method, we choose the following methods as the baselines for the paper recommendation task:

HERec \citep{shi2015semantic} samples nodes using meta-paths and integrates features learned through nonlinear methods to produce the final representation of entities.

PaperRec \citep{zhu2021recommending} utilizes TransD \citep{ji2015knowledge} and Doc2Vec \citep{le2014distributed} to learn features of heterogeneous entity nodes and paper titles, enabling recommendations based on paper similarity.

TAPRec \citep{jiang2023taprec} uses self-attention mechanisms to understand long-term research interests and temporal convolutional networks to identify short-term focuses, combining these insights to provide comprehensive recommendations.

MARec \citep{wang2023marec} employs a GAT-based auxiliary strategy for constructing heterogeneous information networks to enhance feature representation, while utilizing Bi-directional LSTM and an attention mechanism to capture researchers' long-term interests and recent trends.

Besides the specifically designed scientific paper recommendation models, we also compare the following commonly used product recommendation models to investigate the effectiveness of these general-purpose models for recommending scientific papers:

LightGCN \citep{he2020lightgcn} is a simplified GCN for recommendation, which abandons feature transformation and nonlinear activation in traditional GCN.

SimGCL \citep{yu2022graph} employs graph contrastive learning for recommendation which discards the graph augmentations and adds uniform noises to the embedding space for creating contrastive views.

AUPlus \citep{ouyang2024improve} mitigates the negative effects of noisy data by using contrastive learning to improve the alignment and uniformity of user and item representations.

\subsection{Evaluation metrics}
Precision, Recall and normalized Discounted Cumulative Gain (nDCG) are the most commonly used evaluation metrics in the recommendation tasks. Consistent with the previous research, our focus is primarily on the front elements of the recommendation list. Consequently, we employ the Precision@k, Recall@k and nDCG@k metrics, abbreviated as P@k, R@k and N@k, to evaluate the quality of the top-k recommendations. In this paper, k is taken as 5, 10 and 20.

For a target scholar, 
\begin{equation}
P@k = \frac{\left|\hat{P_k}\right| \cap \left|P\right|}{\left|\hat{P_k}\right|},R@k = \frac{\left|\hat{P_k}\right| \cap \left|P\right|}{\left|P\right|},
\end{equation}
where $\hat{P_k}$ is the top-k items of list $\hat{P}$ recommended by the model; $P$ is the ground-truth relevant paper list.

Discounted Cumulative Gain (DCG) evaluates both the quantity and position of correctly recommended items.
\begin{equation}
\mathrm{DCG} @ k=\sum_{i=1}^k \frac{\mathrm{rel}_i}{\log _2(i+1)},
\end{equation}
where $rel_i$ is the Gain list. If the scholar is relevant to the $i$-th item in the top-k recommendation list, then $rel_i$ equals 1; otherwise, it equals 0. nDCG is the normalized DCG,
\begin{equation}
N@k = \frac{DCG_k}{IDCG_k}.
\end{equation}
Here the Ideal Discounted Cumulative Gain (IDCG) represents the DCG of an ideal ordering, with items organized by their relevance coefficients.  

\subsection{Experiment settings}
All the experiments are implemented with Pytorch. The model is optimized using the Adam optimizer and initialized with the Xavier initializer for random parameter initialization. We conduct a grid search to find optimal hyperparameters: batch size of {256, 512, 1024, 2048}, dimensions of {32, 64, 128, 256}, learning rate of {0.1, 0.01, 0.001, 0.0001}, and regularization weight of {0.01, 0.005, 0.001, 0.0005}.

\subsection{Experimental results}
The proposed MIARec model is compared to the baselines to showcase its effectiveness in paper recommendations. We also perform ablation studies to assess the contribution of each module of our model and discuss how parameter variations affect recommendation performance. 

\subsubsection{Model Comparison}
Table \ref{tab:result} presents the model comparison results for P@k, R@k, and N@k (k=5, 10, 20), allowing the following conclusions to be drawn from the results.

\begin{table*}[htbp]
  \centering
  \caption{Performance Comparison of Precision@K, Recall@K, and NDCG@K with Different Ks.}
    \begin{tabular}{cccccccccccc}
    \toprule
    Dataset & \multicolumn{2}{c}{Metric} & HERec & PaperRec & TAPRec & MARec & LightGCN & SimGCL & AUPlus & MIARec & Improv.(\%) \\
    \midrule
    \multirow{9}[6]{*}{DBLP} & \multirow{3}{*}{k=5}  & P@k    & 0.4084 & 0.4372 & 0.5610 & \underline{0.6382} & 0.5474 & 0.5890 & 0.6125 & \textbf{0.6429} & 0.74 \\
                         &   & R@k    & 0.4706 & 0.5068 & 0.5725 & \underline{0.6317} & 0.5748 & 0.6265 & 0.6281 & \textbf{0.6494} & 2.80 \\
                         &   & N@k    & 0.5311 & 0.5482 & 0.5860 & \underline{0.6149} & 0.5749 & 0.5902 & 0.6073 & \textbf{0.6237} & 1.43 \\
                         \addlinespace
                         \cline{2-12}
                         \addlinespace
                         & \multirow{3}{*}{k=10} & P@k    & 0.3729 & 0.4041 & 0.5276 & 0.5831 & 0.5384 & 0.5627 & \underline{0.5902} & \textbf{0.5975} & 1.24 \\
                         &   & R@k    & 0.5002 & 0.5357 & 0.5789 & 0.6471 & 0.6023 & 0.6579 & \underline{0.6624} & \textbf{0.6782} & 2.39 \\
                         &   & N@k    & 0.5520 & 0.5738 & 0.6227 & \underline{0.6213} & 0.5786 & 0.5923 & 0.6188 & \textbf{0.6473} & 4.18 \\
                         \addlinespace
                         \cline{2-12}
                         \addlinespace
                         & \multirow{3}{*}{k=20} & P@k    & 0.3374 & 0.3800 & 0.4872 & \underline{0.5285} & 0.4324 & 0.5193 & 0.5140 & \textbf{0.5304} & 0.36 \\
                         &   & R@k    & 0.5339 & 0.5743 & 0.6187 & 0.6458 & 0.6291 & 0.6620 & \underline{0.6698} & \textbf{0.6873} & 2.61 \\
                         &   & N@k    & 0.5277 & 0.5936 & 0.6116 & \underline{0.6248} & 0.5857 & 0.5983 & 0.6217 & \textbf{0.6492} & 3.91 \\
    \midrule
    \multirow{9}[8]{*}{ACM} & \multirow{3}[0]{*}{K=5} & P@k   & 0.4220  & 0.4498  & 0.5695  & \underline{0.6465}  & 0.5544  & 0.6036  & 0.6237  & \textbf{0.6562 } & 1.50 \\
                        &   & R@k   & 0.4818  & 0.5153  & 0.5827  & \underline{0.6429}  & 0.5826  & 0.6365  & 0.6341  & \textbf{0.6573 } & 2.24 \\
                        &   & N@k   & 0.5388  & 0.5610  & 0.5935  & \underline{0.6213}  & 0.5852  & 0.5953  & 0.6178  & \textbf{0.6301 } & 1.42 \\
                        \addlinespace
                        \cline{2-12}
                        \addlinespace
                        &  \multirow{3}[0]{*}{K=10} & P@k   & 0.3791  & 0.4121  & 0.5407  & 0.5909  & 0.5494  & 0.5736  & \underline{0.5957}  & \textbf{0.6030 } & 1.23 \\
                        &   & R@k   & 0.5119  & 0.5416  & 0.5921  & 0.6536  & 0.6096  & 0.6542  & \underline{0.6716}  & \textbf{0.6881 } & 2.46 \\
                        &   & N@k   & 0.5577  & 0.5867  & 0.6320  & \underline{0.6345}  & 0.5870  & 0.6033  & 0.6255  & \textbf{0.6595 } & 3.94 \\
                        \addlinespace
                        \cline{2-12}
                        \addlinespace
                        &  \multirow{3}[0]{*}{K=20} & P@k   & 0.3464  & 0.3883  & 0.4931  & \underline{0.5403}  & 0.4435  & 0.5319  & 0.5205  & \textbf{0.5463 } & 1.11 \\
                        &   & R@k   & 0.5404  & 0.5829  & 0.6250  & 0.6538  & 0.6406  & 0.6698  & \underline{0.6793}  & \textbf{0.6825 } & 0.47 \\
                        &   & N@k   & 0.5353  & 0.6000  & 0.6215  & 0.6398  & 0.5929  & 0.6108  & \underline{0.6416}  & \textbf{0.6618 } & 3.15 \\
    \bottomrule
    \end{tabular}%
  \label{tab:result}%
\end{table*}%

(1) The proposed MIARec model demonstrates superior performance over baseline approaches across Precision, Recall, and NDCG metrics on both Aminer DBLP and ACM datasets, comprehensively validating its effectiveness. This enhanced performance is primarily attributable to two key factors: First, the model employs a multi-channel aggregation framework that simultaneously learns independent and interdependent embeddings from the heterogeneous academic network. This dual-learning mechanism enables the simultaneous capture of structural similarities within single-relational subgraphs and the latent correlations across diverse scholar relationship types. Second, the introduced mutual-influence-aware embedding methodology systematically models asymmetric academic influence patterns between scholars, thereby optimizing the feature aggregation process during network embedding through context-aware information propagation.

(2) In baseline paper recommendation models, despite all models employing the structural information among scholars, TAPRec and MARec outperform HERec and PaperRec due to their consideration of scholars' interest preferences, showing how academic influence impacts paper selection. However, TAPRec does not utilize graph embedding methods to learn structural information, focusing instead on user preference similarities. Conversely, MARec employs a GAT to learn structural features of heterogeneous networks but falls short in adequately evaluating the mutual influence among scholars. 

(3) The conventional recommendation models - LightGCN, SimgCL, and AUPlus - exhibit inferior performance compared to the MIARec model on this recommendation task. This performance gap is likely due to their lack of a domain-specific design for heterogeneous academic networks, which restricts their capacity to effectively model the complex multi-relational structures characteristic of scholars. Moreover, these methods demonstrate limited exploitation of scholarly content information.

\subsubsection{Ablation study}
We conduct ablation experiments to examine the impact of each module of our MIARec model. 

\begin{figure*}[htb!]
	\centering
	\includegraphics[width=1\textwidth]{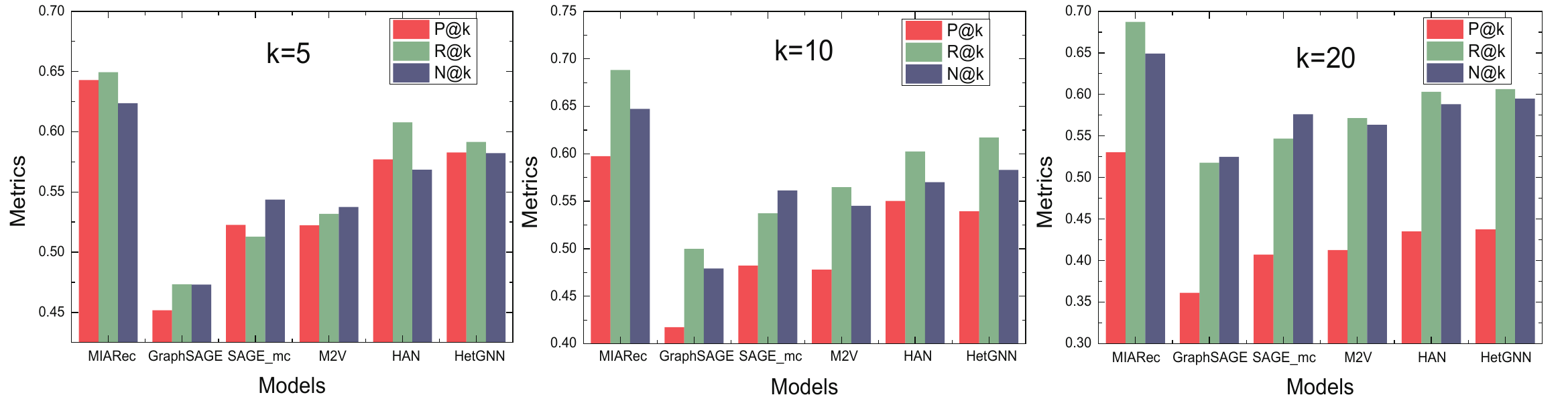}
	\caption{Performance comparison of network embedding methods within MIARec framework. 
    \label{fig:ablation-emb}}
\end{figure*}

\paragraph{Effectiveness of MIA network embedding method}
To ascertain the efficacy of the mutual-influence-aware scholar embedding module within the proposed recommendation system, we replaced this module with the current network embedding models to learn representations of the scholar network, serving as a comparative analysis: GraphSAGE \citep{graphsage} (Since GraphSAGE was originally designed for homogeneous networks, we extend its framework using the proposed multi-channel approach. In each channel, GraphSAGE learns the representation of a specific single-relation sub-network. We refer to this variant method as SAGE\_mc), Metapath2Vec (denote as M2V) \citep{metapath2vec}, HAN \citep{han}, HetGNN \citep{zhang2019heterogeneous}. The comparison result obtained on the DBLP dataset is shown in Figure \ref{fig:ablation-emb}.

\noindent

These comparative models all perform below the MIARec model, indicating that the MIA network embedding module is more capable of effectively acquiring the structural characteristics of scholars required for this task. Among them, GraphSAGE exhibits the lowest accuracy levels across all evaluation metrics, which substantiates the significant limitations of traditional homogeneous graph-based approaches in handling heterogeneous academic network data. However, when employing its enhanced variant, GraphSAGE\_mc, the model performance shows remarkable improvement. This enhancement validates the superior capability of the proposed multi-channel framework in capturing and integrating heterogeneous information in academic networks. The HetGNN model achieves comparatively suboptimal performance, potentially due to its explicit differentiation of diverse edge and node types. Compared to meta-path-based approaches (M2V, HAN), this mechanism preserves a richer original structural information while seamlessly integrating attribute features with structural contexts, thereby enhancing the expressiveness of node representations. However, its inherent limitation lies in the inability to explicitly define semantic relations, which may result in diminished effectiveness when addressing higher-order relationships such as co-venue involving the author-paper-conference triadic structure.

\paragraph{Effectiveness of MIA factor}

The design of the MIA factor $M_{ij}$ in Equation \ref{eq:conv} is crucial for the network embedding module. In Section 4.1.1, we noted that traditional methods using symmetric normalization or the attention mechanism typically set this factor to 1 or the attention weight. Consequently, we define $M_{ij}=1$ and $M_{ij}=att$ in the model, referred to as MIARec\_sn and MIARec\_att, respectively.
\paragraph{Effectiveness of interdependent channel}
The interdependent channel in MIARec captures implicit information across different scholarly relations. To assess its effectiveness, we remove this channel and rely solely on the independent single-relational network embedding channels, referred to as MIARec w/o ic.
\paragraph{Effectiveness of content of papers}
We evaluate the contribution of scientific paper content by removing the paper embedding module from the model, using solely the scholars' representation, referred to as MIARec w/o cont.

\begin{figure*}[htb!]
	\centering
	\includegraphics[width=1\textwidth]{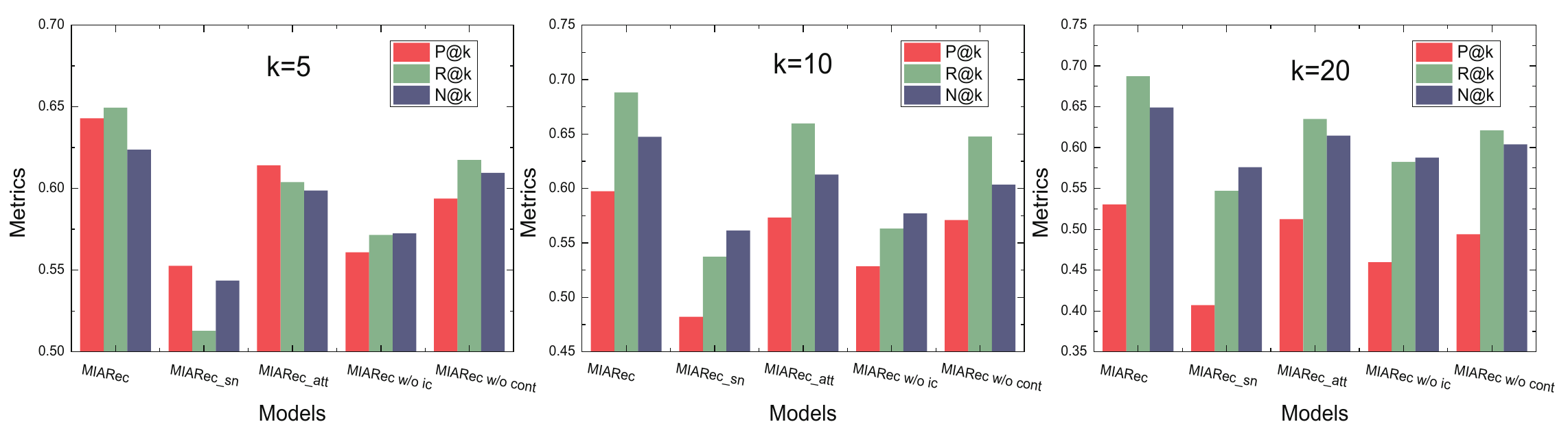}
	\caption{Ablation study of the impact of individual modules on MIARec Performance. 
    \label{fig:ablation}}
\end{figure*}

Figure \ref{fig:ablation} compares MIARec with the above variant models, revealing varying degrees of performance degradation among variants and underscoring the significance of different modules in the MIARec framework. Specifically, the MIARec\_sn variant shows the lowest recommendation accuracy; with parameter M set to 1, it reverts to a conventional GraphSAGE convolutional kernel that aggregates neighboring node features uniformly, neglecting node differences, which severely hampers performance. In contrast, adding an attention mechanism for weight learning during feature aggregation significantly enhances performance, though it still lags behind the comprehensive MIARec model that accounts for the mutual academic influence among scholars.

Removing the interdependent channel significantly reduces model performance, with an average decline of 12.55\%, 15.14\%, 9.50\% across all three metrics. This confirms its contribution to information acquisition beyond independent embedding channels, leading to more accurate network embeddings. Additionally, excluding paper content information results in a performance drop of 6.32\%, 6.81\%, 5.34\%, highlighting the positive impact of paper semantic features on enhancing recommendation outcomes.

\noindent

\begin{figure*}[htb!]
	\centering
	\includegraphics[width=1\textwidth]{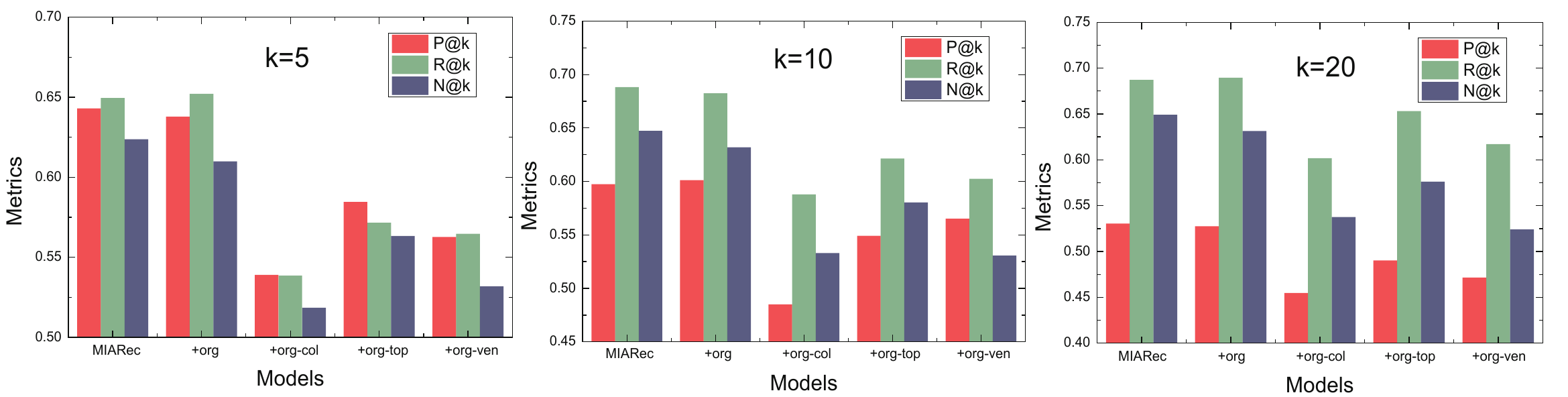}
	\caption{Ablation study of the impact of different scholar relationships on MIARec Performance.\label{fig:ablation-relation}}
\end{figure*}

\paragraph{Effectiveness of scholar relationship selection}
The study also explores how the relationship between scholars in the network embedding module influences outcomes. Noting that shared organizational affiliation is sometimes considered why scholars collaborate. Thus, utilizing the relationship "two scholars belong to the same institution," we extracted a new single-relation sub-network from the academic network and integrated it into the academic network embedding module. To more intuitively highlight the roles of the three selected types of scholar relationships (collaboration, co-topic, co-venue) in the model, we replaced each of these relationships with a co-organization relation network. Consequently, four variant models were constructed, denoted as: +org, +org-col, +org-top, and +org-ven. The comparison results are shown in Figure \ref{fig:ablation-relation}. The results indicate that incorporating the co-organization relation network does not significantly enhance recommendation accuracy across all three evaluation metrics, with some cases even showing a decline. This implies that the co-organization relationship has little impact on scholars' paper preferences and may even hinder recommendation performance by introducing redundant connections. Replacing collaboration, co-topic, and co-venue relationships with co-organization significantly degrades model performance, highlighting their effectiveness in capturing structural similarity among scholars. The worst performance of the +org-col variant further emphasizes the substantial influence of collaboration on scholars' paper preferences.
\noindent

\subsubsection{Hyperparameter selection}
\begin{figure*}[htb!]
	\centering
	\includegraphics[width=1\textwidth]{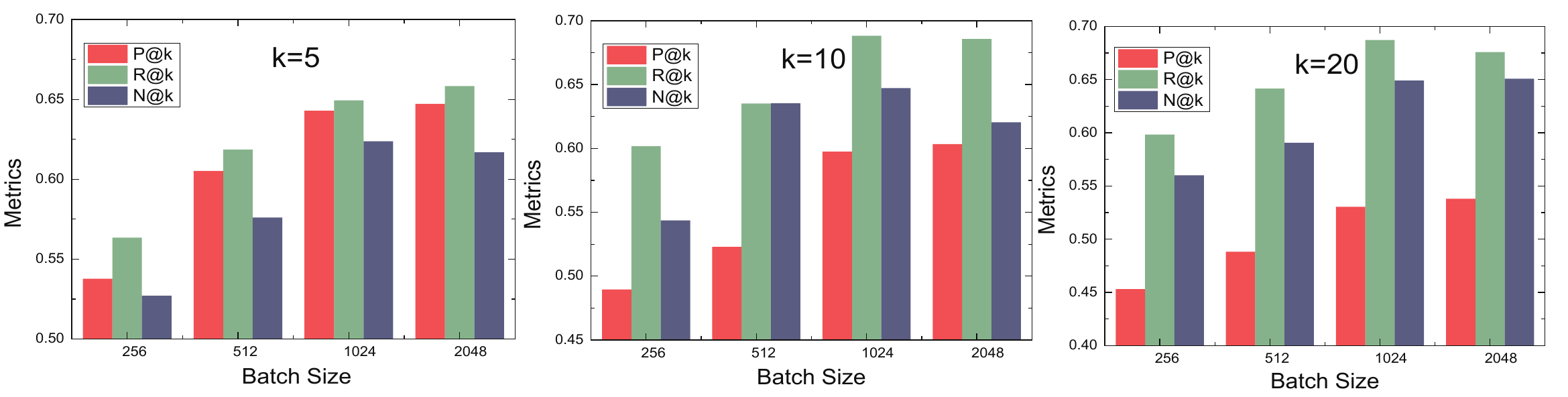}
	\caption{Performance of MIARec models with different batch sizes.\label{fig:batch}}
\end{figure*}

As mentioned in Section 5.4, we utilize grid search to optimize four hyperparameters: batch sizes, dimensions, learning rate, and regularization weight, aiming to identify their optimal values. Figure \ref{fig:batch} illustrates an example of hyperparameter selection, showing that increasing the batch size does lead to improved model performance. However, although a batch size of 2048 slightly outperforms 1024 in overall metrics, the improvement is marginal and even shows a slight decrease in some cases. Moreover, compared to 1024, setting the batch size to 2048 significantly increases space and time costs. Therefore, taking into account the extent of model accuracy improvement and its associated costs, we ultimately chose a batch size of 1024. The selection of the remaining hyperparameters also adheres to the same principles. After considering both performance and resource consumption, the determined optimal parameter settings were as follows: batch size of 1024, dimension of 64, learning rate of 0.001, and regularization weight of 0.0005.

\noindent

\section{Conclusion and future works \label{sec:Conclusion}}
This paper proposes a novel personalized paper recommendation model, MIARec. The model employs a multi-channel approach for heterogeneous network embedding to capture structural features in scholar networks. It divides the network into single-relational sub-networks, learning embeddings for each sub-network independently while identifying interdependent embeddings in a parameter-sharing channel to reveal implicit correlations among scholarly relations. In the embedding process, we introduce a mutual-influence-aware factor to assess the asymmetric academic influence between scholars when aggregating the neighboring features. The resulting scholar embeddings are combined with paper embeddings to enhance recommendation accuracy. Experimental results indicate that our MIARec model outperforms all representative baselines across various metrics, demonstrating the method's effectiveness and stability.

As we have considered the impact of academic influence on the effectiveness of paper recommendations, we see potential in integrating the dynamic characteristics of scholars' academic preferences to better reflect the varying degrees of influence they experience over different periods. Thus, in the future, it is planned to integrate time-aware representations of both academic influence and user preference evolution to advance the performance of scholarly recommendation systems.


\section*{Acknowledgment}
This work is supported by the Chongqing Graduate Research and Innovation Project (No.CYB22128) and the National Natural Science Foundation of China (NSFC) (No.62006198).

\printcredits

%\bibliographystyle{cas-model2-names}
\bibliographystyle{elsarticle-num} %

\bibliography{ref}





% \end{thebibliography}
\end{document}